\documentclass{llncs}

\usepackage{graphicx}
\graphicspath{{./Figures/}}
\usepackage{subfig}
\usepackage{amsmath}
\usepackage{amsfonts}
\usepackage{amssymb}
\usepackage{bm}
\usepackage{mathtools}
\usepackage{nicefrac}
\usepackage{minitoc} 
\usepackage[toc,page]{appendix} 
\usepackage{soul}
\usepackage{esvect}
\usepackage{pgfgantt}
\usepackage{hyperref}
\usepackage{makeidx}
\usepackage[misc]{ifsym}
\usepackage{gensymb}      
\usepackage{hyperref}
\usepackage{widetable}
\usepackage{booktabs}

\begin{document}

\title{Empirical Bayesian Mixture Models \\for Medical Image
Translation} 
\author{Mikael Brudfors\inst{1}  \and John
Ashburner\inst{1} \and Parashkev Nachev\inst{2} \and Ya\"{e}l
Balbastre\inst{1}}

\institute{Wellcome Centre for Human Neuroimaging, UCL, London, UK\\
\email{\{mikael.brudfors.15,j.ashburner,y.balbastre\}@ucl.ac.uk} \and
UCL Institute of Neurology, London, UK\\ \email{p.nachev@ucl.ac.uk}}

\maketitle

\begin{abstract}
Automatically generating one medical imaging modality from another is
known as medical image translation, and has numerous interesting
applications. This paper presents an interpretable generative modelling
approach to medical image translation. By allowing a common model for
group-wise normalisation and segmentation of brain scans to handle
missing data, the model allows for predicting entirely missing
modalities from one, or a few, MR contrasts. Furthermore, the model
can be trained on a fairly small number of subjects. The proposed model
is validated on three clinically relevant scenarios. Results appear
promising and show that a principled, probabilistic model of the
relationship between multi-channel signal intensities can be used to
infer missing modalities -- both MR contrasts and CT images.
\end{abstract}

\section{Introduction}

This paper concerns a relatively simple method of synthesising data of
one medical image modality, from data of other modalities. This is known
as `image translation'. Applications of medical image translation are
numerous, and include \emph{e.g.} harmonising data across scanners;
synthesising computed tomography (CT) images from magnetic resonance
(MR) images for positron emission tomography (PET) attenuation
correction \cite{burgos2013attenuation}, or decrease the need for
radiating a patient; simplifying the problem of multi-modal image
registration \cite{cao2012registration}; or generalising machine
learning techniques by transferring out-of-distribution input data to
the domain of the model's training data \cite{roy2010mr}.

Mapping from the signal intensities of one modality to those of another
can be loosely categorised as either optimisation- or learning-based.
Optimisation-based methods rely only on the data at hand to optimise a
mapping between modalities, and do not use training data. Examples
include using non-parametric joint histograms \cite{kroon2009mri},
estimating an intensity transformation during image registration
\cite{guimond2001three}, and biophysical models
\cite{wein2008automatic}. Learning-based methods use training data to
learn the mapping, and can be applied to translating an unseen image
from one domain into another. Some examples in this category use
clustering \cite{hsu2013investigation}, random forests
\cite{huynh2015estimating}, patch-matching
\cite{iglesias2013synthesizing} and dictionaries
\cite{roy2011compressed}. Learning-based methods based on various
convolutional neural network architectures are currently the most
popular approach for this. Trained end-to-end, on either paired or
unpaired training data
\cite{chartsias2017multimodal,nie2017medical,wolterink2017deep}, they
show promising results at this task, although they can run the risk of
hallucinating unwanted features \cite{cohen2018distribution}.

This paper presents a more interpretable generative modelling approach
to image translation. It could be classed as an optimisation-based
approach, although it does use training data to learn priors that inform
the optimisation of mappings. More specifically, we show how a
generative model for group-wise normalisation and segmentation of
neuroimaging data can be extended to handle missing data. The generative
model has a Gaussian mixture model component, which can naturally handle
missing data \cite{ghahramani1994supervised}. In this paper, we extend
this missing data model to a variational Gaussian mixture. Fitting this
model to various populations of medical images allows us to predict,
from a few MR contrasts, entirely missing modalities (\emph{e.g.},
non-acquired MR contrasts or CT images).

\section{Methods}


The prediction of one modality from another is here cast as a joint
intensity modelling problem. The workhorse of the proposed method is the
unified segmentation model \cite{ashburner2005unified}, which uses
mixtures of Gaussians with non-stationary tissue priors derived from a
deformable template. When a large dataset is available, the optimisation
of the template can be interleaved with the mixture model fit to each
individual subject \cite{blaiotta2018generative}. Furthermore, priors
over the intensity parameters of the Gaussian mixture -- its means and
covariances -- can be defined and optimised as well. This type of
learning, where subject-specific parameters are marginalised while
population parameters are optimised, is known as parametric empirical
Bayesian methods \cite{carlin2000empirical}. Here, exact marginalisation
is intractable, so we resort to a variational approximation.

\subsubsection{Fully observed model:} 

Let $\vec{X} \in \mathbb{R}^{D \times M}$ be a multimodal dataset from
one subject, where $M$ is the number of modalities and $D$ is the number
of voxels in the images. Each voxel is assumed to belong to one of $K$
classes, where the classification is encoded by the label matrix
$\vec{Z} \in \left[0,1\right]^{D\times K}$, with $z_{dk} = 1$
\emph{iff.} voxel $d$ belongs to class $k$. Each tissue class is
associated with a multivariate Gaussian distribution of dimension $M$,
which encodes the intensities' mean ($\boldsymbol{\mu}_{k} \in
\mathbb{R}^M$) and covariance ($\boldsymbol\Sigma_{k} \in \mathbb{R}^{M
\times M}$) over the modalities. The Gaussian mixture model can then be
written as a conditional probability that factorises across voxels:
\begin{align}
p\left(\vec{X}\mid\vec{Z},
\vec{\mu}_{1\dots K}, \vec{\Sigma}_{1\dots K}\right)
= \prod_{d=1}^D \prod_{k=1}^K 
\mathcal{N}\left(\vec{x}_{d}\mid
\vec{\mu}_{k},\vec{\Sigma}_{k}\right)^{z_{dk}}.
\end{align}

Subject-specific parameters (the label matrix and Gaussian parameters)
are assumed to be drawn from prior distributions that describe their
variability at the population level. Labels are drawn from a categorical
distribution whose probabilities are encoded by a deformable template
$\vec{a} \in \mathbb{R}^{D_a \times K}$. This template is mapped to the
subject's brain using a non-linear deformation field $\vec{\phi}$. This
assumption can be written as the conditional likelihood:
\begin{align}
p\left(\vec{Z}\right)
= \prod_{d=1}^D 
\operatorname{Cat}\left(\vec{z}_{d}\mid\vec{\pi}_{d}\right),
~~
\vec{\pi}_d  \in \mathbb{R}^K, ~~ \pi_{dk} = 
\frac{\exp(\omega_k + \vec{a}_{dk}(\vec{\phi}))}{\sum_{j=1}^K 
\exp(\omega_{j} + \vec{a}_{dj}(\vec{\phi}))}, 
\end{align}
where $\vec{\omega} \in \mathbb{R}^K$ is a vector of global class
proportions, which can be optimised to account for variable amounts of
different classes (an example when modelling brain images could be
atrophy due to ageing). The Gaussian parameters are drawn from their
conjugate Gauss-Wishart distribution:
\begin{align}
p\left(\vec{\mu}_{k}, \vec{\Sigma}_{k}^{-1}\right)
={} &
\mathcal{NW}\left(\vec{\mu}_{k}, \vec{\Sigma}_{k}^{-1}
\mid \vec{\mu}_{0k}, b_{0k},
\vec{V}_{0k}, \nu_{0k}\right)\cr
={} & \mathcal{N}\left(\vec{\mu}_{k} \mid \vec{\mu}_{0k}, 
\vec{\Sigma}_{k} / b_{0k}\right)
\mathcal{W}\left(\vec{\Sigma}_{k}^{-1} \mid \vec{V}_{0k}, 
\nu_{0k}\right)
.
\end{align}

Assuming that all population parameters are fixed, a mean-field
approximation is made so that the posterior distribution over all
latent, subject-specific parameters factorises as:
\begin{align}
q\left(\vec{Z},\vec{\mu}_{1\dots K},
\vec{\Sigma}_{1\dots K}^{-1}\right)
=
\left[\prod_{d=1}^D
q\left(\vec{z}_d\right)
\right]
\left[
\prod_{k=1}^K
q\left(\vec{\mu}_k,\vec{\Sigma}^{-1}_k\right)
\right],
\end{align}
with $q\left(\vec{z}_d\right)
=\operatorname{Cat}\left(\vec{z}_d\mid\tilde{\vec{z}}_d\right)$ and
$q\left(\vec{\mu}_k,\vec{\Sigma}^{-1}_k\right)
=\mathcal{NW}\left(\vec{\mu}_k,\vec{\Sigma}^{-1}_k\mid
\tilde{\vec{\mu}}_{k}, \tilde{b}_{k}, \tilde{\vec{V}}_{k},
\tilde{\nu}_{k}\right)$. The posterior parameters (denoted by a tilde)
can be optimised in turn by maximising the evidence lower bound (ELBO):
\begin{align}
\mathcal{L} ={} & 
\mathbb{E}\left[\ln p\left(\vec{X}\mid
\vec{Z},\vec{\mu}_{1\dots K}, 
\vec{\Sigma}_{1\dots K}\right)\right] \cr
&
-\sum_{d=1}^D\operatorname{D}_{\mathrm{KL}}\left(q_{\vec{z}_d}~\middle\|~p_{\vec{z}_d}\right)
-\sum_{k=1}^K\operatorname{D}_{\mathrm{KL}}\left(q_{\vec{\mu}_k,\vec{\Sigma}_k}~\middle\|~p_{\vec{\mu}_k,\vec{\Sigma}_k}\right).
\end{align}
When multiple subjects $\left\{\vec{X}_n\right\}_{n=1}^N$ are processed,
the posterior distribution factorises across subjects and a combined
ELBO can be written by summing the individual ELBOs ($\mathcal{L} =
\sum_{n=1}^N \mathcal{L}_n$). In this case, empirical population priors
can be obtained by optimising the combined ELBO with respect to the
template ($\vec{a}$) and Gauss-Wishart prior hyper-parameters
($\vec{\mu}_{0k}, b_{0k}, \vec{V}_{0k}, \nu_{0k}$). The means and scale
matrices have closed form solutions, while the template and degrees of
freedom must be optimised using an iterative scheme. Population prior
parameters and subject posterior parameters can be optimised in turn,
resulting in a variational Expectation-Maximisation (VEM) algorithm
\cite{blaiotta2016variational}.

\subsubsection{Missing modalities:}

Let us assume that some modalities are missing in a voxel\footnote{For
example, a multi-channel MRI might have three contrasts: T1w, T2w and
PDw. In one voxel, only the T1w intensity is observed. The T2w and PDw
intensities are then assumed missing in that voxel. Note that different
voxels can have different combinations of contrasts/modalities
missing.}. We write as $\vec{o}$ the vector indexing observed modalities
and as $\vec{m}$ the vector indexing missing modalities. Therefore, the
observed channels can be written as $\vec{g}=\vec{x}_{\vec{o}}$ and the
missing channels as $\vec{h}=\vec{x}_{\vec{m}}$, where the voxel index
$d$ has been temporarily dropped for clarity. For a voxel in class $k$,
the marginal distribution of the observed channels can then be written
as \cite{bishop2006pattern}:
\begin{align}
p\left(\vec{g}\mid \vec{\mu}_k,\vec{\Sigma}_k, z_k=1\right)
= \mathcal{N}\left(\vec{g}\mid\vec{\mu}_{k\vec{o}}, 
\vec{\Sigma}_{k\vec{o}\vec{o}}\right),
\end{align}
and the conditional distribution of the missing channels as:
\begin{align}
&p\left(\vec{h}\mid \vec{g}, 
\vec{\mu}_k,\vec{\Sigma}_k, z_k=1\right)
= \cr
&\mathcal{N}\left(\vec{h}\mid
\vec{\mu}_{k\vec{m}} - 
\left(\vec{\Lambda}_{k\vec{m}\vec{m}}\right)^{-1}\vec{\Lambda}_{k\vec{m}\vec{o}}\left(\vec{g}
- \vec{\mu}_{k\vec{o}}\right),
\left(\vec{\Lambda}_{k\vec{m}\vec{m}}\right)^{-1}
\right),
\end{align}
where the precision matrix $\vec\Lambda=\vec\Sigma^{-1}$ is the inverse
of the covariance matrix.

The set of all missing values in an image is written as $\mathcal{H} =
\left\{\vec{h}_d\right\}_{d=1}^D$. The mean field approximation becomes:
\begin{align}
q\left(\mathcal{H},\vec{Z},\vec{\mu}_{1\dots K},
\vec{\Sigma}^{-1}_{1\dots K}\right)
=
\left[\prod_{d=1}^D
q\left(\vec{h}_d\mid\vec{z}_d\right)
q\left(\vec{z}_d\right)
\right]
\left[
\prod_{k=1}^K
q\left(\vec{\mu}_k,\vec{\Sigma}^{-1}_k\right)
\right],
\end{align}
where $q\left(\vec{h}_d\mid\vec{z}_d\right) = \prod_{k=1}^K
\mathcal{N}\left(\vec{h}_d\mid\tilde{\vec{h}}_{dk},
\tilde{\vec{S}}_{dk}\right)^{z_{dk}}$. The marginal posterior over
missing values is a mixture of Gaussians that can be obtained by
marginalising
the labels: 
\begin{align}
q(\vec{h}_d)= \sum_{k=1}^K \tilde{z}_{dk}\mathcal{N}
\left(\vec{h}_d\mid\tilde{\vec{h}}_{dk},\tilde{\vec{S}}_{dk}\right).
\label{eq:predmiss}
\end{align}
Its expected value is $\mathbb{E}\left[\vec{h}_d\right] = \sum_k
\tilde{z}_{dk}\tilde{\vec{h}}_{dk}$. This is the expression that we
evaluate to predict missing voxels.

The set of all observed values is written as $\mathcal{G} =
\left\{\vec{g}_d\right\}_{d=1}^D$. The ELBO can then be written in two
equivalent forms:
\begin{align}
\mathcal{L} ={} & 
\mathbb{E}\left[\ln p\left(\mathcal{G}\mid
\vec{Z},\vec{\mu}_{1\dots K},
\vec{\Sigma}_{1\dots K}\right)\right] \cr
&-\sum_{d=1}^D\operatorname{D}_{\mathrm{KL}}\left(q_{\vec{z}_d}~\middle\|~p_{\vec{z}_d}\right)
-\sum_{k=1}^K\operatorname{D}_{\mathrm{KL}}\left(q_{\vec{\mu}_k,\vec{\Sigma}_k}~\middle\|~p_{\vec{\mu}_k,\vec{\Sigma}_k}\right)
\label{eq:elbo1} \\
\mathcal{L} ={} & 
\mathbb{E}\left[\ln p\left(\vec{X}\mid
\vec{Z},\vec{\mu}_{1\dots K},
\vec{\Sigma}_{1\dots K}\right)\right]
-\sum_{d=1}^D\mathbb{E}_{\vec{z}_d}\left[\operatorname{D}_{\mathrm{KL}}\left(q_{\vec{h}_d\mid\vec{z}_d}~\middle\|~p_{\vec{h}_d\mid\vec{z}_d}\right)\right]
\nonumber\\ &
-\sum_{d=1}^D\operatorname{D}_{\mathrm{KL}}\left(q_{\vec{z}_d}~\middle\|~p_{\vec{z}_d}\right)
-\sum_{k=1}^K\operatorname{D}_{\mathrm{KL}}\left(q_{\vec{\mu}_k,\vec{\Sigma}_k}~\middle\|~p_{\vec{\mu}_k,\vec{\Sigma}_k}\right).
\label{eq:elbo2}
\end{align}
The first form is used to optimise the labels' posterior parameters,
while the second is used to optimise, in turn, the missing values and
the Gaussian posterior parameters.

\subsubsection{Model updates:}

Optimising the ELBOs in \eqref{eq:elbo1} and \eqref{eq:elbo2} gives the
subject-level update equations as:
\begin{align}
\tilde{z}_{dk} ={} & 
\frac{\exp\left(\mathbb{E}\left[\ln\mathcal{N}\left(\vec{g}_d \mid 
\vec{\mu}_k,\vec{\Sigma}_k\right)\right]+\ln\pi_{dk}\right)}{\sum_{l=1}^K
\exp\left(\mathbb{E}\left[\ln\mathcal{N}\left(\vec{g}_d \mid 
\vec{\mu}_l,\vec{\Sigma}_l\right)\right]+\ln\pi_{dl}\right)}.
\\
\tilde{b}_{k} ={} & b_{0k} + \sum_{d=1}^D \tilde{z}_{dk} 
\\
\tilde{\vec{\mu}}_{k} ={} & 
\frac{b_{0k}\vec{\mu}_{0k} + \sum_{d=1}^D 
\mathbb{E}\left[z_{dk}\vec{x}_d\right]}{\tilde{b}_{k}}
\\
\tilde{\nu}_{k} ={} & \nu_{0k} + \sum_{d=1}^D 
\tilde{z}_{dk} 
\\
\tilde{\vec{V}}_{k}^{-1} ={} & 
\nu_{0k}\vec{V}_{0k}^{-1} +  \sum_{d=1}^D 
\mathbb{E}\left[z_{dk}\vec{x}_d\vec{x}_d^{\mathrm{T}}\right]
+ b_{k0}\vec{\mu}_{k0}\vec{\mu}_{k0}^{\mathrm{T}}
- 
\tilde{b}_{k}\tilde{\vec{\mu}}_{k}\tilde{\vec{\mu}}_{k}^{\mathrm{T}}
\end{align}
The update equations for the Gaussian parameters in the missing data
case are very similar to the fully observed case, except that
expectations are taken about the data. These expectations are evaluated
as:
\begin{equation}
\begin{aligned}[c]
\mathbb{E}\left[z_{dk}\vec{x}_d\right]_{\vec{o}} ={} & 
\tilde{z}_{dk}\vec{g}_d,
\\
\mathbb{E}\left[z_{dk}\vec{x}_d\right]_{\vec{m}} ={} & 
\tilde{z}_{dk}\tilde{\vec{h}}_{dk},
\\
\mathbb{E}\left[z_{dk}\vec{x}_d\vec{x}_d^{\mathrm{T}}\right]_{\vec{o}\vec{o}}
={} & 
\tilde{z}_{dk}\vec{g}_d\vec{g}_d^{\mathrm{T}},
\end{aligned}
\qquad
\begin{aligned}[c]
\mathbb{E}\left[z_{dk}\vec{x}_d\vec{x}_d^{\mathrm{T}}\right]_{\vec{m}\vec{m}}
={} & 
\tilde{z}_{dk}\left(\tilde{\vec{h}}_{dk}\tilde{\vec{h}}_{dk}^{\mathrm{T}}
+\tilde{\vec{S}}_{dk}\right),
\\
\mathbb{E}\left[z_{dk}\vec{x}_d\vec{x}_d^{\mathrm{T}}\right]_{\vec{o}\vec{m}}
={} & 
\tilde{z}_{dk}\vec{g}_d\tilde{\vec{h}}_{dk}^{\mathrm{T}},
\\
\mathbb{E}\left[z_{dk}\vec{x}_d\vec{x}_d^{\mathrm{T}}\right]_{\vec{m}\vec{o}}
={} & 
\tilde{z}_{dk}\tilde{\vec{h}}_{dk}\vec{g}_d^{\mathrm{T}},
\end{aligned}
\end{equation}
where
\begin{equation}
\begin{aligned}[c]
\tilde{\vec{h}}_{dk} ={} &
\tilde{\vec{\mu}}_{k\vec{m}} 
- 
\tilde{\vec{\Lambda}}_{k\vec{m}\vec{m}}^{-1}\tilde{\vec{\Lambda}}_{k\vec{m}\vec{o}}
\left(\vec{g}_d - \tilde{\vec{\mu}}_{k\vec{o}}\right),
\end{aligned}
\qquad
\begin{aligned}[c]
\tilde{\vec{S}}_{dk} ={} &
\tilde{\vec{\Lambda}}_{k\vec{m}\vec{m}}^{-1},
\end{aligned}
\end{equation}
and $\tilde{\vec{\Lambda}}_k = \tilde{\nu}_k\tilde{\vec{V}}_k$ is the
posterior expected precision matrix of a given class.

Finally, we provide the optimal updates of the Gaussian prior
parameters, given a set of individual posterior parameters. All prior
parameters have closed-form updates, except for the degrees of freedom
of the Wishart distribution, which is updated using an iterative
Gauss-Newton scheme. The update equations are:
\begin{align}
\vec{\mu}_{0k} ={} & 
\left(\sum_{n=1}^N 
\tilde{\nu}_{nk}\tilde{\vec{V}}_{nk}\right)^{-1}
\left(\sum_{n=1}^N 
\tilde{\nu}_{nk}\tilde{\vec{V}}_{nk}\tilde{\vec{\mu}}_{nk}\right),
\\
b_{0k}^{-1} ={} & \frac{1}{NM} \sum_{n=1}^N 
\tilde{\nu}_{nk}
\left(\vec{\mu}_{0k}-\tilde{\vec{\mu}}_{nk}\right)^{\mathrm{T}}
\tilde{\vec{V}}_{nk}
\left(\vec{\mu}_{0k}-\tilde{\vec{\mu}}_{nk}\right),
\\
\vec{V}_{0k} ={} & 
\frac{1}{N\nu_{0k}}\sum_{n=1}^N 
\tilde{\nu}_{nk}\tilde{\vec{V}}_{nk},
\\
\frac{\partial\mathcal{L}}{\partial \nu_{0k}} ={} &
-\frac{1}{2}\left(
N\left(\ln\left|\vec{V}_{0k}\right| 
+ \psi_M\left(\frac{\nu_{0k}}{2}\right)\right)
- \sum_{n=1}^N \left(\ln\left|\tilde{\vec{V}}_{nk}\right| 
- \psi_M\left(\frac{\tilde{\nu}_{nk}}{2}\right)\right)
\right),
\\
\frac{\partial^2\mathcal{L}}{\partial \nu_{0k}^2} ={} &
-\frac{N}{4}\psi'_M\left(\frac{\tilde{\nu}_{0k}}{2}\right).
\end{align}
We do not provide update rules for the template ($\vec{a}$), as they can
be found in \cite{blaiotta2018generative}.

\section{Experiments and Results}

In this section we aim to explore the translation (or inference)
capability of the proposed model by conducting three experiments on
publicly available data. We investigate: (1) inferring missing voxels of
MRIs with differing field of views; (2) inferring entirely missing MRI
contrasts; and (3), inferring CT scans from MRIs. The findings are
quantified by computing the peak-signal-to-noise-ratio (PSNR) for an
image channel $c$ as:
\begin{align}
\text{PSNR} = 10\log_{10}\frac{maxval^2}{MSE},
\end{align}
where the mean-squared error is defined as $\text{MSE} =
\frac{1}{D}\sum_{d=1}^D (\hat{x}_{cd} - (\mathbb{E}\left[
\vec{h}_{d}\right])_c)^2$, $maxval$ is the maximum channel intensity in
the reference image $\hat{\vec{X}}$, and $\mathbb{E}\left[
\vec{h}_{d}\right]$ from \eqref{eq:predmiss} is evaluated to predict
missing voxels. The PSNR is a metric that is commonly used in the
medical image synthesis literature
\cite{chartsias2017multimodal,wolterink2017deep,nie2017medical}. Note
that no voxels are excluded when computing the PSNR.

\begin{figure*}[t]
\centering
\includegraphics[width=\textwidth]{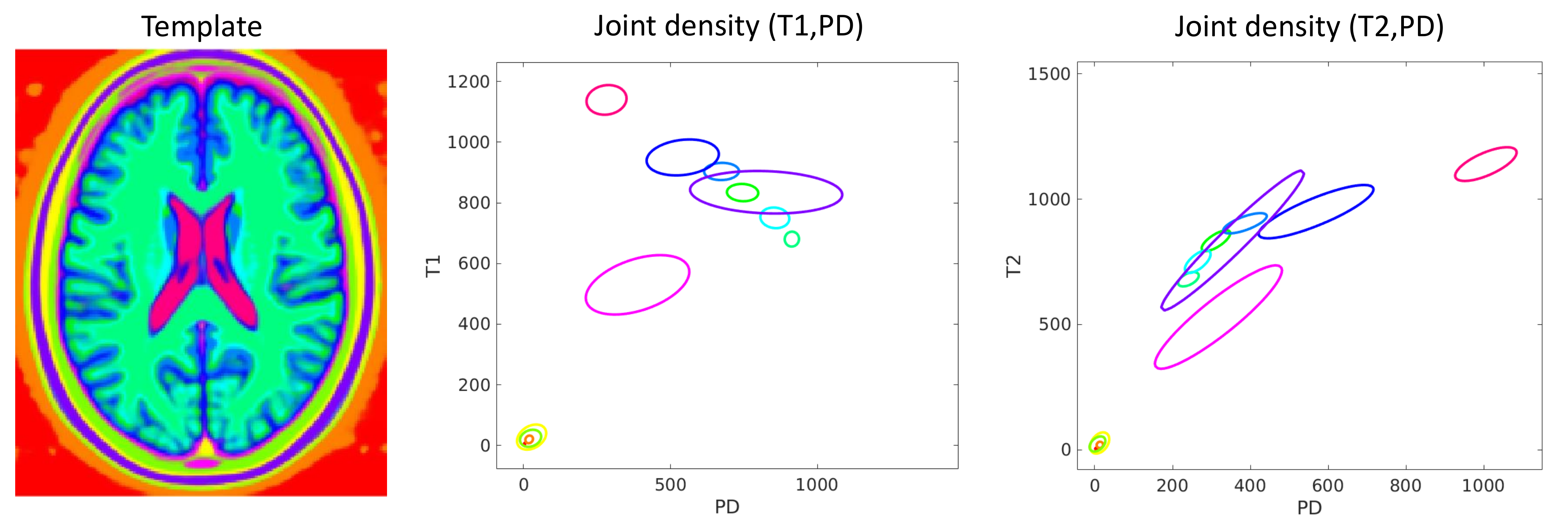}
\caption{Template and expectations of the Gaussians drawn from the
Gauss-Wishart priors, learnt from 50 IXI subjects. Densities are plotted
using their $3\sigma$ isocontours. This model is fit to a new subject,
which allows for inferring missing voxels.}
\label{fig:iximodel}
\end{figure*}

\subsection{MRI Contrast Translation}

This section evaluates translating between MR contrasts. The model is
trained on 50 subjects from the publicly available IXI
dataset\footnote{\url{http://brain-development.org/ixi-dataset/}}, which
was acquired on three different MR scanners\footnote{This scenario is
more realistic in a clinical context. The results would improve if data
from only one scanners was used.}. Each IXI subject has three MR images:
a T1-, T2- and PD-weighted scan (T1w, T2w and PDw). Furthermore, the
images have approximately 1 mm isotropic voxels and all subjects are
healthy. $K=12$ mixture components are used, resulting in the model
shown in Fig. \ref{fig:iximodel}. Note that the template learned by the
algorithm does not need to represent real tissues.  Here, the model has
been treated as a method of representing a probability density function,
rather than as a way to do clustering. Any `meaningful' clusters are
incidental.

\begin{figure*}[t]
\centering
\includegraphics[width=\textwidth]{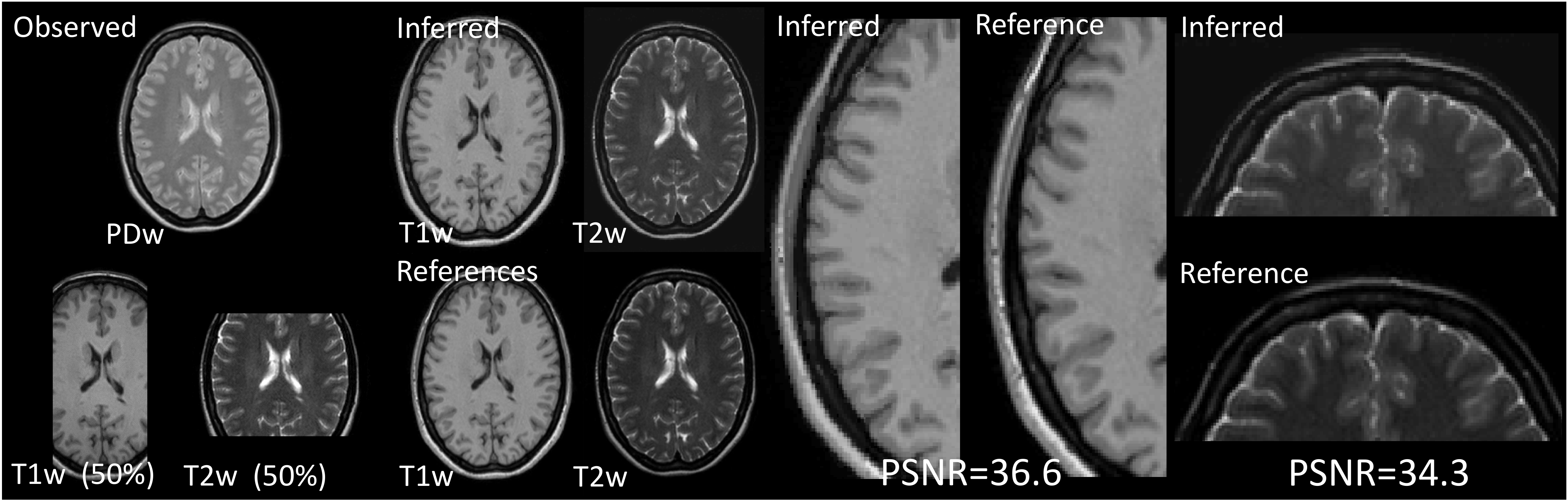}
\caption{Example of inferring MRIs with differing field of views. An MR
image with three channels (PDw, T1w and T2w) is observed. The PDw scan
has full brain coverage, while the T1w and T2w scans have partial brain
coverage (50\% of voxels removed in each channel). From the observed
data the values of the missing T1w and T2w voxels are inferred. The
reference T1w and T2w scans are shown for comparison, as well as PSNR
values.}
\label{fig:fov}
\end{figure*}

\subsubsection{Inferring MRIs with Differing Fields of View:}

Doctors often acquire routine clinical MR scans of multiple contrasts.
Commonly, these contrasts have differing fields of view, meaning the
brain coverage varies (\emph{cf.} observed T1w and T2w images in Fig.
\ref{fig:fov}). This can be problematic for image segmentation routines
as voxels with non-observed contrasts need to be discarded. The model
should prevent this issue by inferring the values of these missing
voxels. To test this, T1w, T2w and PDw scans of 50 unseen IXI subjects
are used\footnote{The model is trained on IXI subjects
\texttt{IXI[064-118]}, and tested on \texttt{IXI[002-063]}.}. All of the
voxels are retained in the PDw image, while an increasing amount of
voxels are removed from the T1w and T2w images (25\%, 50\%, 75\% and
100\%). The missing voxels are then inferred with the trained model. An
example can be seen in Fig. \ref{fig:fov}. The mean PSNR computed
between the known references and the inferred images are shown in Table
\ref{fig:fov}. For routine clinical MRI, it is rare that more than 50\%
of the field of view is missing. The results therefore suggest that the
model does a good job at filling in missing fields of view, which could
be of value in segmenting hospital data.

\begin{table*}
\fontsize{8}{7.2}\selectfont
\centering
\caption{Results for inferring MR images with different fields of view
(for 50 subjects). The PSNR is computed between known T2w and PDw
references and inferred images, where an increasing percentage of the
field of view has been removed. Results are shown as mean$\pm$std. }
\begin{widetable}{\columnwidth}{c | c c c c} \toprule
& 
\multicolumn{4}{c}{\textbf{PSNR}} 
\\
\textbf{Contrast} & \textbf{25\%} & \textbf{50\%} & 
\textbf{75\%} & \textbf{100\%} 
\\\midrule
T1w & $42.1\pm$1.6 & $36.3\pm1.3$ & $31.1\pm1.3$ & 
$28.9\pm1.2$  \\
T2w & $40.7\pm$2.1 & $34.4\pm2.0$ & $30.4\pm1.8$ & $27.6\pm1.6$ 
\\\bottomrule
\end{widetable}
\label{tab:fov}
\end{table*}

\begin{figure*}[t]
\centering
\includegraphics[width=0.5\textwidth]{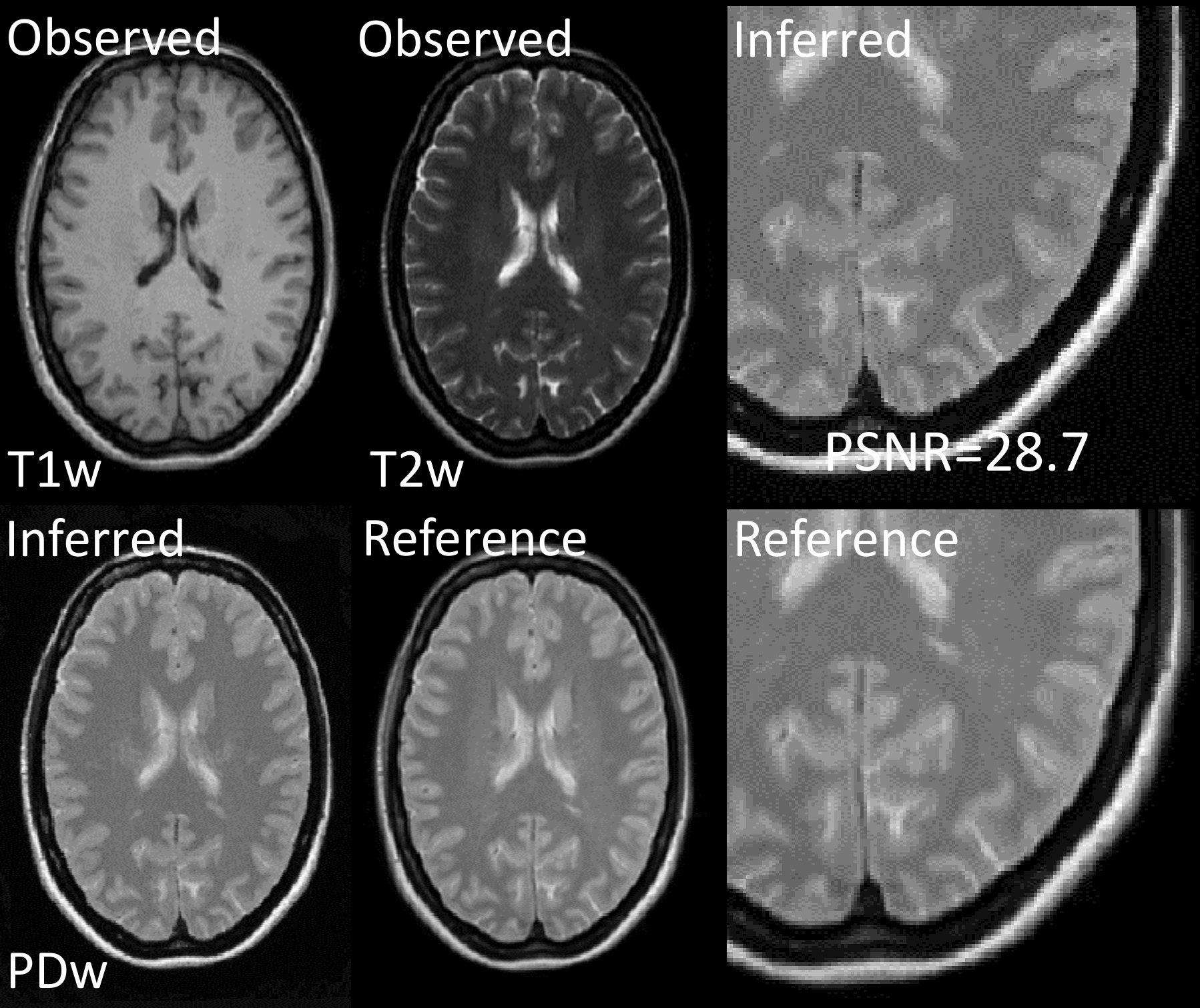}
\caption{Example of inferring non-acquired MR contrasts. An MR image
with two channels (T1w and T2w) is observed. The PDw scan is missing,
but inferred from the observed T1w and T2w scans. The reference PDw scan
is shown for comparison, as well as the PSNR value.}
\label{fig:chn}
\end{figure*}

\subsubsection{Inferring MR Contrasts:}

Could the proposed model be used to infer an entirely missing MR
contrast? An interesting application for this type of MRI translation
could be for segmentation methods based on deep learning. A deep
learning model that has been trained on MR images of a specific contrast
can overfit to its training data \cite{brudfors2019nonlinear}. If images
could be simulated as to match the training data of the deep learning
model, it might generalise better.

To test how well the model predict a missing contrast the same IXI
subjects as in the previous experiment are used. For each subject, all
combinations of contrasts are permuted over, set as either observed or
missing. For example, we observe just the T1w image and infer the T2w
and PDw scans, or we observe the T2w and PDw scans and infer the T1w
(see Fig. \ref{fig:chn}). The results from this experiment are shown in
Table \ref{tab:chn}. These results imply that the T1w image is the most
predictive, as the lowest PSNR is obtained when this contrast is
missing. The example inferred PDw image in Fig. \ref{fig:chn} looks
realistic when compared to the known reference, although more noisy. The
results in Table \ref{tab:chn} are close to those previously reported in
the literature \cite{chartsias2017multimodal} (for the same task but a 
different dataset).

\begin{table*}
\fontsize{8}{7.2}\selectfont
\centering
\caption{Results for inferring MR image contrasts (for 50 subjects).
PSNR is computed for all different permutations of observed and missing
contrasts. Results are shown as mean$\pm$std.}
\begin{widetable}{\columnwidth}{c c | c c c} \toprule
\multicolumn{2}{c|}{\textbf{Contrasts}} & 
\multicolumn{3}{c}{\textbf{PSNR}} \\
\textbf{Observed} & \textbf{Missing} & \textbf{T1w} & 
\textbf{T2w} & 
\textbf{PDw} 
\\\midrule
T1w & T2w, PDw & - & $28.9\pm1.5$ & $28.5\pm1.1$ \\ 
T2w & T1w, PDw & $28.2\pm1.0$ & - & $28.3\pm1.5$ \\
PDw & T1w, T2w & $28.0\pm1.2$ & $27.6\pm1.6$ & - \\
T1w,T2w & PDw & - & - & $28.8\pm0.9$ \\
T2w,PDw & T1w & $29.2\pm1.4$ & - & -  \\
T1w,PDw & T2w & - & $28.1\pm1.5$ & -
\\\bottomrule
\end{widetable}
\label{tab:chn}
\end{table*}

\begin{figure*}[t]
\centering
\includegraphics[width=0.85\textwidth]{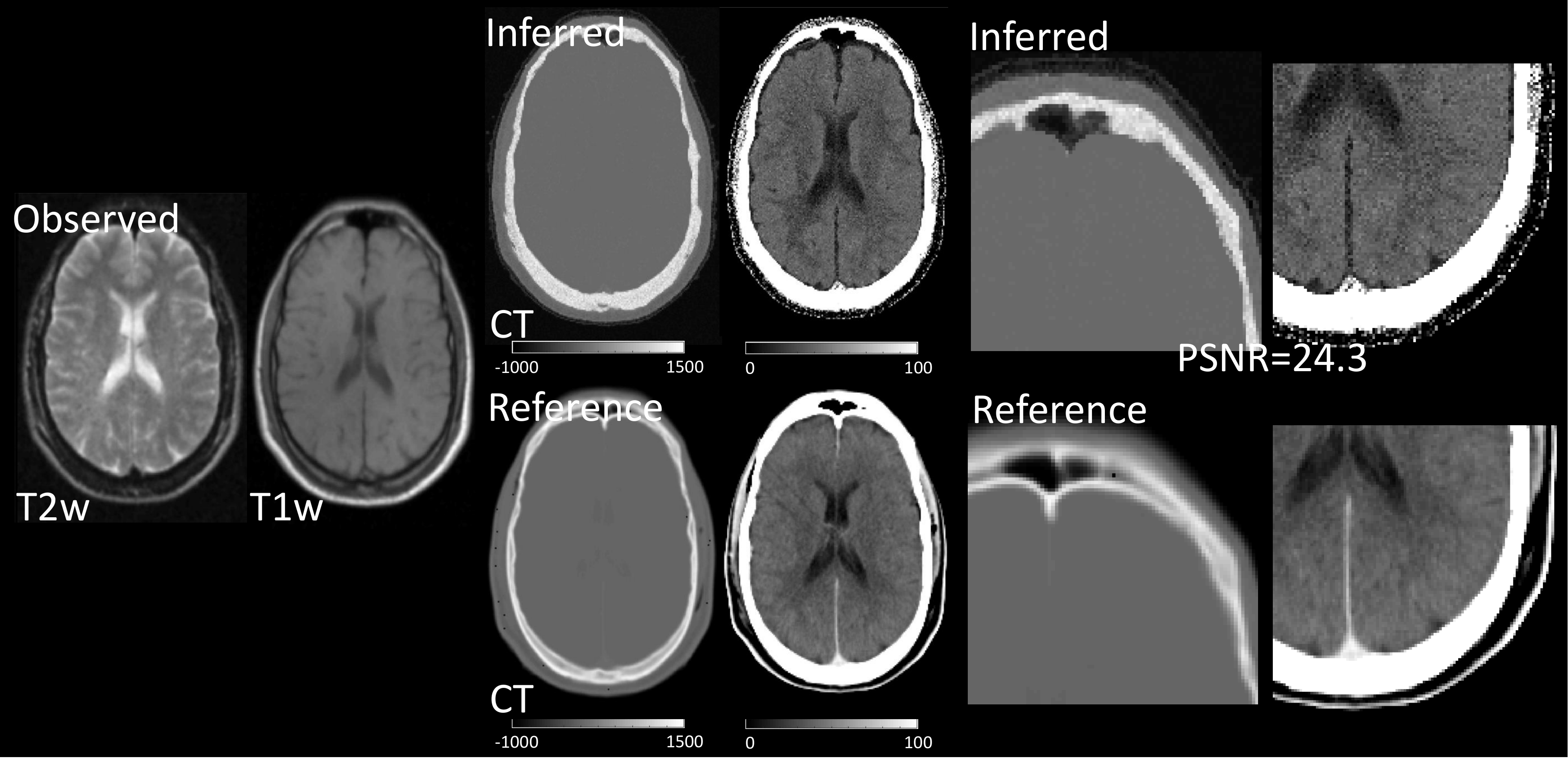}
\caption{Example of MRI to CT translation. An MR image with two channels
(T1w and T2w) is observed. A CT scan is then inferred from the observed
T1w and T2w scans. The reference CT scan is shown for comparison, as
well as the PSNR value.}
\label{fig:ct}
\end{figure*}

\subsection{MRI to CT Translation}

Accurately translating MRIs to CTs is interesting for numerous reasons,
\emph{e.g.}, for removing the exposure to radiation that CT imaging
involves, or for attenuation correction in MR-PET imaging. The proposed
model should allow for this type of translation, by training it on
subjects who have both MR and CT imaging. We therefore retrain the
intensity distribution hyper-parameters of the model -- retaining the
template learnt from the IXI dataset -- on eight patients from the RIRE
dataset\footnote{\url{https://www.insight-journal.org/rire/}}
\cite{west1996comparison}. Each patient in this dataset contains a
number of imaging modalities. Here, only the patients with T1w and T2w
MR scans (non-rectified), and CT images, are used. Note that the RIRE
dataset is challenging to use due to the images having thick-slices,
sometimes pathology, as well as requiring an initial co-registration
(the dataset is part of a registration challenge and therefore
purposefully misaligned). Each subject's scans are registered using the
co-registration routine of the SPM12 software.

To test the models ability to translate MRIs to CTs, eight unseen RIRE
patients are used\footnote{The model is trained on RIRE patients
\texttt{patient[102-109]}, and tested on
\texttt{patient[001-007,101]}.}. The trained model is fit to each
subject's T1w and T2w scans. The expected marginal posterior
distribution over the missing CT image can then be computed. An example
is shown in Fig. \ref{fig:ct}. The mean$\pm$std PSNR between the
inferred CT images and the known references is $25.5\pm1.2$. Considered
the intensity hyper-parameters were trained on only eight subjects, the
results are satisfactory, although not on pair with deep learning based
techniques \cite{wolterink2017deep}. The examples images in Fig.
\ref{fig:ct} suggests that the model does not capture a detailed enough
distribution of bone. Additionally, the meninges does not appear in the
inferred image, but is instead modelled as cerebrospinal fluid. Fitting
not only the intensity hyper-parameters to the CT data, but also the
template, could resolve these issues. More training data would also
help.



\section{Conclusion}

This paper showed how a popular model for segmenting brain scans -- a
probabilistic forward model with a Gaussian mixture part -- can be
extended to infer missing data. For multi-channel segmentation, this
extension circumvents the need to model only voxels that are observed in
all channels. It furthermore enables predicting one MR contrast from
another, or CTs from MRIs. The model gives reasonable results if trained
on a small number of subjects, but we would expect further improvements
with access to more training data. Interestingly, image translation is
just a `by-product' of learning the parameters of a joint probability
distribution that models missing voxels. The same model can also be used
to segment, bias correct and spatially normalise brain scans.

The model requires setting the number of Gaussian mixture components
($K$) at the start of the training. If this number is set too low, then
the simulated images will look unrealistic. Here, this issue was
resolved by using a fairly large number of components, which was found
empirically capturing a detailed enough model distribution.
Uninformative mixture components can then be drived to zero, due to the
Bayesian setting of the Gaussian mixture model, by making point
estimates of the values of the global tissue proportions
($\vec{\omega}$). This is known as automatic relevance determination
\cite{bishop2006pattern}.

Generative modelling approaches integrating multi-channel images, like
the one presented here, should involve a component that relates signal
across the various channels. The approach presented in this paper
involves a probabilistic model of the relationship between signal
intensities over channels. An alternative approach would be to use a
multi-channel total variation (MTV) prior, which ensures that `edges'
appear in similar locations across channels. The MTV prior can be used
to achieve super-resolution or denoising of medical images
\cite{brudfors2018mri}. An avenue of future work could therefore be to
incorporate both of these components into a super-resolution method, to
improve resolution of thick-sliced, hospital-grade MR scans. By
combining, for example, axial thick-sliced T2-weighted images and
sagittal thick-sliced T1w images of the same subjects. In this example,
the T2w image could provide some of the missing T1w signal in the
left-right direction, whereas the T1w image could fill in some of the
missing T2w signal in the inferior-posterior direction. Of course, this
strategy would need to be formulated properly, but this work aimed to
show a proof of the concept that one of those components, a
probabilistic model between channels, does a good job at filling in
missing data in MR images.

%
%

\subsubsection*{Acknowledgements:} MB was funded by the EPSRC-funded
UCL Centre for Doctoral Training in Medical Imaging (EP/L016478/1) and
the Department of Health’s NIHR-funded Biomedical Research Centre at
University College London Hospitals. MB and JA was funded by the EU
Human Brain Project's Grant Agreement No 785907 (SGA2). YB was funded
by the MRC and Spinal Research Charity through the ERA-NET Neuron joint
call (MR/R000050/1).

\bibliography{bibliography}

\begin{thebibliography}{10}

\bibitem{burgos2013attenuation}
N.~Burgos, M.~J. Cardoso, M.~Modat, S.~Pedemonte, J.~Dickson, A.~Barnes, J.~S.
  Duncan, D.~Atkinson, S.~R. Arridge, B.~F. Hutton, {\em et~al.}, ``Attenuation
  correction synthesis for hybrid {PET-MR} scanners,'' in {\em MICCAI},
  pp.~147--154, Springer, 2013.

\bibitem{cao2012registration}
T.~Cao, C.~Zach, S.~Modla, D.~Powell, K.~Czymmek, and M.~Niethammer,
  ``Registration for correlative microscopy using image analogies,'' in {\em
  WBIR}, pp.~296--306, Springer, 2012.

\bibitem{roy2010mr}
S.~Roy, A.~Carass, N.~Shiee, D.~L. Pham, and J.~L. Prince, ``{MR} contrast
  synthesis for lesion segmentation,'' in {\em ISBI}, pp.~932--935, IEEE, 2010.

\bibitem{kroon2009mri}
D.-J. Kroon and C.~H. Slump, ``{MRI} modalitiy transformation in demon
  registration,'' in {\em ISBI}, pp.~963--966, IEEE, 2009.

\bibitem{guimond2001three}
A.~Guimond, A.~Roche, N.~Ayache, and J.~Meunier, ``Three-dimensional multimodal
  brain warping using the demons algorithm and adaptive intensity
  corrections,'' {\em IEEE T. Med. Imaging}, vol.~20, no.~1, pp.~58--69, 2001.

\bibitem{wein2008automatic}
W.~Wein, S.~Brunke, A.~Khamene, M.~R. Callstrom, and N.~Navab, ``Automatic
  {CT}-ultrasound registration for diagnostic imaging and image-guided
  intervention,'' {\em Med. Image Anal.}, vol.~12, no.~5, pp.~577--585, 2008.

\bibitem{hsu2013investigation}
S.-H. Hsu, Y.~Cao, K.~Huang, M.~Feng, and J.~M. Balter, ``Investigation of a
  method for generating synthetic {CT} models from {MRI} scans of the head and
  neck for radiation therapy,'' {\em Phys. Med. Biol.}, vol.~58, no.~23,
  p.~8419, 2013.

\bibitem{huynh2015estimating}
T.~Huynh, Y.~Gao, J.~Kang, L.~Wang, P.~Zhang, J.~Lian, and D.~Shen,
  ``Estimating {CT} image from {MRI} data using structured random forest and
  auto-context model,'' {\em IEEE T. Med. Imaging}, vol.~35, no.~1,
  pp.~174--183, 2015.

\bibitem{iglesias2013synthesizing}
J.~E. Iglesias, E.~Konukoglu, D.~Zikic, B.~Glocker, K.~Van~Leemput, and
  B.~Fischl, ``Is synthesizing {MRI} contrast useful for inter-modality
  analysis?,'' in {\em MICCAI}, pp.~631--638, Springer, 2013.

\bibitem{roy2011compressed}
S.~Roy, A.~Carass, and J.~Prince, ``A compressed sensing approach for {MR}
  tissue contrast synthesis,'' in {\em IPMI}, pp.~371--383, Springer, 2011.

\bibitem{chartsias2017multimodal}
A.~Chartsias, T.~Joyce, M.~V. Giuffrida, and S.~A. Tsaftaris, ``Multimodal mr
  synthesis via modality-invariant latent representation,'' {\em IEEE
  transactions on medical imaging}, vol.~37, no.~3, pp.~803--814, 2017.

\bibitem{nie2017medical}
D.~Nie, R.~Trullo, J.~Lian, C.~Petitjean, and M.~Ruan, ``Medical image
  synthesis with context-aware generative adversarial networks,'' pp.~417--425,
  Springer, 2017.

\bibitem{wolterink2017deep}
J.~M. Wolterink, A.~M. Dinkla, M.~H. Savenije, P.~R. Seevinck, C.~A. van~den
  Berg, and I.~I{\v{s}}gum, ``Deep {MR} to {CT} synthesis using unpaired
  data,'' in {\em SASHIMI}, pp.~14--23, Springer, 2017.

\bibitem{cohen2018distribution}
J.~P. Cohen, M.~Luck, and S.~Honari, ``Distribution matching losses can
  hallucinate features in medical image translation,'' in {\em MICCAI},
  pp.~529--536, Springer, 2018.

\bibitem{ghahramani1994supervised}
Z.~Ghahramani and M.~I. Jordan, ``Supervised learning from incomplete data via
  an {EM} approach,'' in {\em NeurIPS}, pp.~120--127, 1994.

\bibitem{ashburner2005unified}
J.~Ashburner and K.~J. Friston, ``Unified segmentation,'' {\em NeuroImage},
  vol.~26, no.~3, pp.~839--851, 2005.

\bibitem{blaiotta2018generative}
C.~Blaiotta, P.~Freund, M.~J. Cardoso, and J.~Ashburner, ``Generative
  diffeomorphic modelling of large {MRI} data sets for probabilistic template
  construction,'' {\em NeuroImage}, vol.~166, pp.~117--134, 2018.

\bibitem{carlin2000empirical}
B.~P. Carlin and T.~A. Louis, ``Empirical {Bayes}: {Past}, present and
  future,'' {\em J. Am. Stat. Assoc.}, vol.~95, no.~452, pp.~1286--1289, 2000.

\bibitem{blaiotta2016variational}
C.~Blaiotta, M.~J. Cardoso, and J.~Ashburner, ``Variational inference for
  medical image segmentation,'' {\em Comput. Vis. Image Und.}, vol.~151,
  pp.~14--28, 2016.

\bibitem{bishop2006pattern}
C.~M. Bishop, {\em Pattern recognition and machine learning}.
\newblock Springer, 2006.

\bibitem{brudfors2019nonlinear}
M.~Brudfors, Y.~Balbastre, and J.~Ashburner, ``Nonlinear markov random fields
  learned via backpropagation,'' in {\em IPMI}, pp.~805--817, Springer, 2019.

\bibitem{west1996comparison}
J.~B. West, $\dots$, and R.~P. Woods, ``Comparison and evaluation of
  retrospective intermodality image registration techniques,'' in {\em Medical
  {{Imaging}} 1996: {{Image Processing}}}, vol.~2710, pp.~332--348, {SPIE},
  1996.

\bibitem{brudfors2018mri}
M.~Brudfors, Y.~Balbastre, P.~Nachev, and J.~Ashburner, ``{MRI}
  super-resolution using multi-channel total variation,'' in {\em MIUA},
  pp.~217--228, Springer, 2018.

\end{thebibliography}
\bibliographystyle{ieeetr}

\end{document}